# Characteristic structure of star-forming clouds

**Philip C. Myers**

Harvard-Smithsonian Center for Astrophysics, 60 Garden Street,

Cambridge MA 02138 USA

pmyers@cfa.harvard.edu

Abstract. This paper gives a new way to diagnose the star-forming potential of a molecular cloud region from the probability density function of its column density (*N*-pdf). It gives expressions for the column density and mass profiles of a symmetric filament having the same *N*-pdf as a filamentary region. The central concentration of this characteristic filament can distinguish regions and can quantify their fertility for star formation. Profiles are calculated for *N*-pdfs which are pure lognormal, pure power law, or a combination. In relation to models of singular polytropic cylinders, characteristic filaments can be unbound, bound, or collapsing depending on their central concentration. Such filamentary models of the dynamical state of *N*-pdf gas are more relevant to star-forming regions than are models of spherical collapse. The star formation fertility of a bound or collapsing filament is quantified by its mean mass accretion rate when in radial free fall. For a given mass per length, the fertility increases with the filament mean column density and with its initial concentration. In selected regions the fertility of their characteristic filaments increases with the level of star formation.

# 1. Introduction

The structural properties of nearby star-forming molecular clouds have received increased attention recently, as parsec-scale maps of dust column density have become available, based mainly on star-count near-infrared extinction (Lada et al. 1994, Alves et al. 1998, Lombardi et al. 2014), and on far-infrared emission observed with the *Herschel Space Observatory* (Pilbratt et al. 2010, André & Saraceno 2005).

These data have revealed two key properties: cloud images are remarkably filamentary on a wide range of scales (André et al 2010, Molinari et al. 2010, Arzoumanian 2011), and the probability density functions of their column density maps ($N$-pdfs) have particular shapes. Generally clouds with little star formation have a $N$-pdf which is primarily lognormal (LN), while clouds with substantial star formation have a $N$-pdf which is LN at low density and has a power-law (PL) tail at high density (LNPL; Kainulainen et al. 2009, Schneider et al. 2014*a*). Some regions forming massive stars have LNPL $N$-pdfs which are dominated by their PL component (Lombardi et al. 2014), and some $N$-pdfs change from LNPL to pure PL as the region over which the $N$-pdf is evaluated is reduced from a giant molecular cloud (GMC) to a smaller, denser, infrared dark cloud (IRDC) embedded in the GMC (Schneider et al. 2014*b*).

Uncertainties associated with low column density measurement, with foreground and background column density, and with the choice of cloud boundary make the LN portion of some $N$-pdfs poorly determined (Lombardi et al. 2015). In contrast, simulations of supersonic turbulent gas clearly indicate LN shapes at low density, with or without self-gravity, as well as PL tails at high density when self-gravity is included (Vazquez-Semadeni 1994, Ballesteros-Paredes et al. 2011, Kritsuk et al. 2011).

Formation of filamentary structure is generally ascribed to supersonic colliding flows, a consequence of HD or MHD turbulent motions (Mac Low & Klessen 2004), or to interacting shocks (Pudritz & Kevlahan 2013). Filamentary structures which resemble observed clouds are seen in simulations of supersonic turbulence (Padoan et al. 2001), of turbulence and self-gravity (Heitsch et al. 2008, Gomez & Vazquez-Semadeni 2014), and of turbulence, self-gravity and magnetic fields (Li et al. 2010, Chen & Ostriker 2014). Filaments which are self-gravitating become more concentrated and persist longer than unbound filaments.

Similarly, the tendency for $N$-pdfs and $n$-pdfs to have LN shape is also ascribed to interactions of turbulent flows and shocks (Vazquez-Semadeni 1994, Ballesteros-Paredes et al. 2011). The transition from LN to a LNPL or PL shape is attributed to the development of regions which are self-gravitating (Vazquez-Semadeni 1994, Klessen 2000) and which are collapsing globally, locally, or both (Ballesteres-Paredes et al. 2011, Federrath & Klessen 2013). Thus filamentary structure and PL tails of $N$-pdfs are both associated with the development of dense self-gravitating gas in turbulent regions.

Most interpretation of the PL property of $N$-pdfs has focussed on spherically symmetric collapse. The density profile of a sphere in free fall collapse from rest approaches a power law with log-log slope -12/7 (Penston 1969). The profile of a collapsing singular isothermal sphere, including thermal pressure, approaches -3/2 (Shu 1977). These density profile power laws translate to $N$-pdf power-law slopes - 2.2 to -2.3 (Federrath & Klessen 2013). An ensemble of uniform spheres collapsing in free fall leads to a power-law $N$-pdf slope of -2.1 (Girichidis et al. 2014). This range of "spherical collapse slopes" -2.1 to -2.3 includes several of the slopes estimated from extinction studies of nearby dark clouds (Kainulainen et al. 2009).

However, interpretation of PL $N$-pdfs in terms of spherically collapsing gas faces several difficulties. First, most of the gas which contributes to the PL slope of a typical $N$-pdf consists of elongated filaments rather than clumps or cores which could be approximated as spherical. This dominance of filaments is evident in recent *Herschel* studies having well-resolved map structure and well-determined $N$-pdfs, in Orion (Stutz & Kainulainen 2015) and Cyg X (Schneider et al. 2015). In Aquila, the gas in the PL portion of the $N$-pdf has more than 50% of its mass in filaments but only ~ 15% in prestellar cores (Andre 2014, Könyves et al. 2015).

Second, observed $N$-pdf PL slopes appear to have no statistically significant preference for spherical collapse values. In *Herschel* studies reporting well-determined PL slopes, 11 clouds with a variety of star formation activity have slopes from -1.3 to -4.1, but none of these slopes lies in the above spherical collapse range -2.1 to -2.3 (Schneider et al. 2013, 2014a, b; 2015). In the Perseus molecular cloud, slopes vary from -1.4 to -2.9 in seven clumps, with two in the spherical collapse range (Sadavoy et al. 2014). In the Orion A cloud, slopes vary from -0.93 to -3.0 in eight regions, with two in the spherical collapse range (Stutz & Kainulainen 2015). The number of regions having spherical collapse slope (four) is indistinguishable from that due to random sampling of a uniform distribution of slopes in each sample.

Third, if all the gas in a typical PL *N*-pdf were in star-forming collapse, the resulting star formation rate would be unrealistically high. In the Aquila complex described above, the total gas mass in the PL *N*-pdf is approximately $2 \times 10^3$ $M_\odot$ (Könyves et al. 2015). If this gas were in star-forming collapse with mass efficiency 0.1, typical of cores and clumps in Oph and Per (Jorgensen et al. 2008), and if it were forming stars having the mean mass of the initial mass function, 0.36 $M_\odot$ (Wiedner & Kroupa 2006), some 560 protostars would result in the next few free-fall times. This number is greater by a factor ~10 than the ~50 Class 0 protostars currently known in the same region (Bontemps et al. 2010). Such Class 0 protostars have probably formed in the last few free-fall times, or in the last few 0.1 Myr (Dunham & Vorobyov 2012). Such an tenfold increase in the star formation rate, from the last few 0.1 Myr to the next few 0.1 Myr, seems highly unlikely.

These points suggest that the PL gas in the typical *N*-pdf refers mostly to filamentary gas, and that star-forming cores are included only at the very highest column densities. The wide range of observed PL slopes suggests that the filamentary gas does not have a unique radial structure arising from a single dynamical state, such as pressure-free collapse. Instead a range of radial concentrations of filamentary gas would seem to be required for consistency. Therefore this paper investigates how *N*-pdfs relate to the radial structure of their filamentary gas.

To relate a *N*-pdf to filament structure, the change of variables relation between probability density functions is used to find the projected structure of a symmetric filament having a given *N*-pdf. This "inversion" of a *N*-pdf is more general than in earlier studies, because it does not need to assume particular column density structures and test the resulting *N*-pdfs against an observed *N*-pdf by trial and error. Instead it assumes only that the column density decreases monotonically with projected radius, and it finds the symmetric column density profile whose *N*-pdf matches the given *N*-pdf exactly.

For many PL *N*-pdfs, the inferred projected structure is similar to that of a singular polytropic cylinder, a self-gravitating cylinder whose radial density profile varies as $n \sim r^{-p}$, with $1 \leq p < 2$ (Toci & Galli 2015; TG15). This structure is that of a regular polytropic cylinder having the same exponent *p*, in the limit of infinite central density or infinite radius. This structure is also identical to that of a "Plummer-like" cylinder with the same *p*, in the same limit of infinite central density or radius (Fischera 2014). The unbound, bound, and collapsing properties of such power-law cylinders provide a useful framework for interpreting the dynamical state and star formation fertility of *N*-pdf gas.

Section 2 describes the restrictions, assumptions and results linking an observed *N*-pdf to a characteristic spatial structure. Section 3 compares these results with models of filamentary cloud structure and Section 4 applies them to observed *N*-pdfs. Section 5 gives a discussion, and Section 6 summarizes the paper.

## 2. Inversion of a *N*-pdf to obtain a characteristic structure

2.1. Requirements and properties

The log of column density $N$ with respect to its mean value $\overline{N}$ over a map is called $\eta \equiv \ln(N/\overline{N})$, and the probability that $\ln(N/\overline{N})$ lies between $\eta$ and $\eta + d\eta$ is called $p(\eta)\,d\eta$, where $p(\eta)$ is the probability density (Federrath & Klessen 2013). Here $p(\eta)$ is denoted *N*-pdf as in Rathborne et al. (2014), to avoid confusion with the *n*-pdf, the pdf of log volume density.

The models here rely on column density data based on near-infrared extinction of background stars by dust (Lombardi & Alves 2001, Lombardi 2009), and on far-infrared dust emission (Schneider et al. 2012, 2014a). Each method is more uncertain at low column density (Lombardi et al. 2015), and in many cases it is difficult to distinguish "cloud" gas from "background" gas for visual extinction $A_V$ < ~ 1 magnitude.

The column density $N$ of the characteristic condensation is assumed to vary smoothly and monotonically with one projected space coordinate $b$. This assumption allows use of the change of variables relation between two probability density functions. This relation is based on the fact that the probability contained in a differential area is invariant under a change of variables. For condensations assumed to be centrally concentrated, $N$ decreases as $b$ increases from zero to a finite maximum $R$. An estimate of the typical radial extent $R$, of order 1 pc, is given in Section 2.5.

Then since $\eta$ increases with $N$ and projected area $\alpha$ increases with $b$,

$$p(\eta)d\eta = -p(\alpha)d\alpha \quad . \qquad (1)$$

Although the focus of this paper is on filamentary structure, the inversion analysis also applies to projected structures having circular symmetry. Here the inversion is described for both kinds of symmetry, and from Section 2.2 onward the discussion is limited to filaments.

In equation (1), the enclosed area is $a = \pi b^2$ for projected condensations having circular symmetry, such as idealized spherical models of "cores" and "clumps." Similarly, $a = 2Lb$ for projected condensations having linear symmetry with respect to a midline of projected length $L$. Such linear-symmetric condensations include filaments whose 3D shapes may be cylindrical, flattened, or more complex.

The condensation derived in this way is denoted a characteristic condensation, and its structure is called a characteristic structure. The procedure is called "inversion" because it contrasts with "forward" calculations where an underlying structure is assumed in order to derive the corresponding $N$-pdf (Press et al. 2007). In this procedure the calculation goes in the reverse direction, from $N$-pdf to underlying structure.

If a region is dominated by one or a few condensations with similar structure, or if many similar condensations occupy a large enough fraction of the region area, the derived characteristic structure may also be representative of the condensations. The characteristic structure may therefore be useful to analyze and compare regions in different clouds, or in different parts of the same cloud. In principle, the large number of measurements in a region can provide a representative structure with better sensitivity and less statistical fluctuation than those of a single structure observed in the same region. Limitations on useful application of this inversion are discussed in Section 5.1.

The characteristic structure discussed here is derived mostly from observations having typical spatial resolution of a few 0.01 pc in regions within a few hundred pc of the Sun (Kainulainen et al. 2009, Arzoumanian et al. 2011, Schneider et al. 2014*a*). It does not take into account more detailed structure inferred from analysis of spectral lines. Some line observations suggest that filaments inferred from dust continuum maps are better described as bundles of nearly thermal "fibers" (Hacar & Tafalla 2011, Hacar et al. 2013, Tafalla & Hacar 2014). If so, the typical continuum structure may be understood as a statistical average over finer-scale structure.

The following analysis proceeds on the assumption that the conditions for useful inversion of $N$-pdfs of continuum column density apply sufficiently well.

Assuming that all map data points have the same area resolution, the right-hand side of equation (1) equals $-2\pi b db/(\pi R^2)$ for circular symmetry, or $-2L db/(2LR)$ for linear symmetry, giving

$$p(\eta)d\eta = -d(\beta^D) \qquad (2)$$

where $\beta \equiv b/R$ is the normalized projected radius, and $D = 1$ or $2$ for linear or circular symmetry. Then integration of equation (2) gives

$$\int_{\eta(l)}^{\eta(u)} d\eta\, p(\eta) = \int_{\beta(l)}^{\beta(u)} d(\beta^D) \qquad (3)$$

where each integral is taken from a lower value $(l)$ to an upper value $(u)$, where $\beta(l)$ corresponds to $\eta(u)$ and where $\beta(u)$ corresponds to $\eta(l)$.

Equation (3) gives the basic relation between column density and normalized projected coordinate, from inversion of the $N$-pdf. Equation (3) applies to any $N$-pdf which is continuous and properly normalized. The pdf need not have an analytic form. Then the definition of $\eta$ gives the column density profile of the characteristic structure as

$$N_{pdf}(\beta) = \overline{N} \exp[\eta(\beta)] . \qquad (4)$$

A relation similar to equation (4) also exists between the volume density $n$ and the normalized spatial coordinate, derived from inversion of the $n$-pdf. However since the $n$-pdf is not directly observable, this relation is not discussed further here.

The following sections 2.2-2.4 give expressions for the analytic inversion of LN, LNPL, and PL *N*-pdfs to obtain profiles of column density and enclosed mass. The reader interested primarily in results may find it more convenient to proceed to Section 2.5.

2.2. Inversion of a lognormal (LN) *N*-pdf

2.2.1. LN *N*-pdf model

In some clouds with relatively little star formation, the *N*-pdf is well approximated by a normal function, as in Lupus V, LDN 1719, and the California Nebula (Kainulainen et al. 2009), in the well-studied Galactic Center cloud G0.253+0.016 (the "brick," Johnston et al. 2014, Rathborne et al. 2014), and in the "quiet" region of the Polaris Cloud (Schneider et al. 2014*a*). The corresponding pdf of $N/\overline{N}$ is therefore well approximated by a LN function. In such regions, the *N*-pdf can be written

$$p(\eta) = \frac{1}{\sqrt{2\pi}\sigma_\eta} \exp(-u^2) \qquad (5)$$

where

$$u \equiv \frac{\eta - \eta_0}{\sqrt{2}\sigma_\eta} \qquad (6)$$

and where $\eta_0$ and $\sigma_\eta$ are respectively the mean and standard deviation of the normal distribution of $\eta$.

2.2.2. LN column density profile

Equations (3) and (4) with $\eta(l) = -\infty$, $\eta(u) = \eta$, $\beta(l) = 0$, and $\beta(u) = \beta$ give

$$\eta(\beta) = \eta_0 + \sqrt{2}\sigma_\eta \, \text{erf}^{-1}(1 - 2\beta^D) \qquad (7)$$

where $\text{erf}^{-1}(x)$ is the inverse error function of $x$. The mean of the log column density $\eta_0$ is obtained from the means of the normal and lognormal probability functions and from the requirement that the mean of $N$ over its distribution be $\overline{N}$, giving $\eta_0 = -\sigma_\eta^2$ (Vazquez-Semadeni 1994, Ostriker et al. 2001). Equation (7) shows that for each symmetry dimension $D = 1$ or 2, $\eta(\beta)$ depends only on the width parameter $\sigma_\eta$.

The LN column density profile is obtained from equations (4-7). It declines steeply at small and large radius, and declines more gradually at mid-range radii, for both circular and linear symmetry. **Figure 1** shows a LN $N$-pdf having width parameter $\sigma_\eta = 0.5$ typical of observed clouds, and **Figure 2** shows the corresponding LN column density profiles $N_{\text{pdf}}(\beta)$ for $D = 1$ and $D = 2$. The profile declines more steeply with radius for $D = 2$ than for $D = 1$.

The LN column density profile $N_{\text{pdf}}(\beta)$ is centrally infinite, as for power laws of $\beta$, but it does not decline as an exact power law of $\beta$. Instead its log-log slope decreases slowly for small $\beta$. The inverse error function $\text{erf}^{-1}(1-2\beta^D)$ in equation (7) can be approximated as $(-\ln 4\beta^D)^{1/2}$ for $\beta \ll 1$ (Winitzki 2003), whence the log-log slope for $D = 1$ is $-0.43\ \sigma_\eta/\sqrt{2}$ for $\beta = 0.001$ and $-0.56\ \sigma_\eta/\sqrt{2}$ for $\beta = 0.01$. These slopes (-0.20 and -0.15 respectively) are close to zero for the typical width $\sigma_\eta = 0.5$, indicating that the typical LN column density profile at small radius is only slightly more condensed than a uniform column density profile.

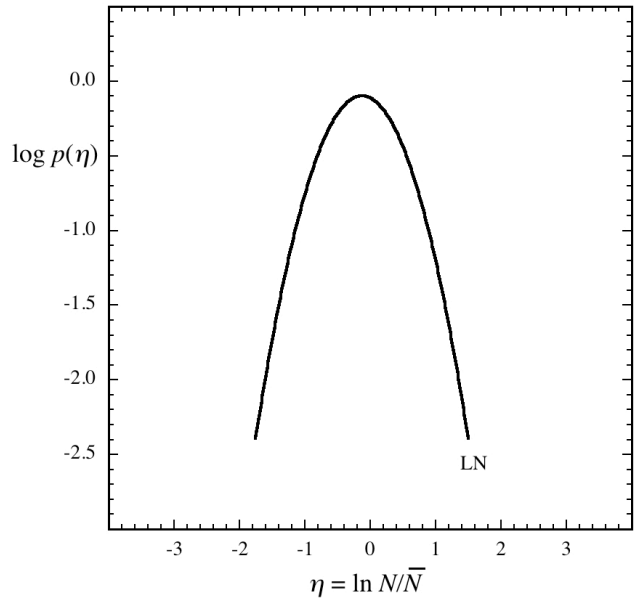

**Figure 1.** Lognormal $N$-pdf having width parameter $\sigma_\eta = 0.5$.

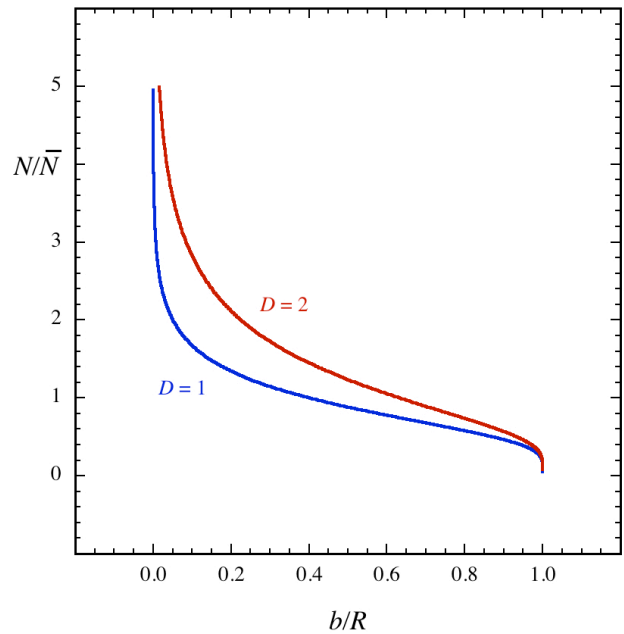

**Figure 2.** Log column density $N$ as a function of projected radius $b$, for projected condensations with circular symmetry ($D = 2$) or linear symmetry ($D = 1$) characteristic of a lognormal $N$-pdf having width parameter $\sigma_\eta = 0.5$ as in Figure 1. The column density is normalized to its mean value $\overline{N}$ and the projected radius is normalized to its maximum value $R$.

The extreme values of the LN column density profiles at small radius correspond to the highest column density values in the cloud maps. These values tend to have large statistical error since they are based on few data points. Furthermore, the typical presence of two or three central fibers in spectral-line studies of nearby filaments (Hacar & Tafalla 2011) challenges the assumption that column density declines monotonically from a single maximum. Thus the detailed filament structure $N_{pdf}(\beta)$ derived from $N$-pdf inversion is likely to be uncertain at its smallest radii.

The mass profile is a more robust measure of filament structure than the column density, since it is an integrated quantity and is less sensitive to central fine structure. As an example, consider two cylinders with maximum radius $R$, a singular cylinder with $n = n_0(r/a)^{-3/2}$ and a Plummer cylinder with $n = n_0[1+(r/a)^2]^{-3/4}$, where $n_0$ is a constant and $a$ is a thermal scale length. Each has the same large-scale density profile $n \sim r^{-3/2}$ as in the LNPL model of Section 2.3.3. Their central column densities have infinite ratio, but their half-mass radii have fractional difference $\sim 2(a/R)^{1/2}$ when $R >> a$. For $R = 1$ pc and $a = 0.01$ pc, their half-mass radii differ by $\sim 20\%$. Thus for comparisons of observations and models it may be more useful to compare enclosed mass than column density.

2.2.3. LN enclosed mass profile

The mass $M$ enclosed within coordinate $\beta$ is obtained by integrating the column density profile over its coordinate extent from the midline for $D = 1$, or over the enclosed area when $D = 2$, up to the maximum mass $M_{max}$ when $\beta = 1$. For $D = 1$, $M_{max} = 2m\overline{N}RL$ where $m$ is the mean molecular mass and $L$ is the filament length. The mass per length is then $\Lambda = M/L$ and its maximum value is $\Lambda_{max} = M_{max}/L$. For $D = 2$, the maximum mass is $M_{max} = \pi m\overline{N}R^2$. The enclosed mass normalized to its maximum value is denoted $\lambda_{LN}(\beta)$, and is given by

$$\lambda_{LN}(\beta) = \frac{1}{2}\left[1 + \mathrm{erf}\left(\frac{\sigma_\eta}{\sqrt{2}} - \mathrm{erf}^{-1}(1-2\beta^D)\right)\right] \quad . \tag{8}$$

The enclosed mass fraction as a function of radius, given in equation (8), increases from zero to its maximum value of unity, as shown in **Figure 3**. The mass increases more rapidly for $D = 2$ than for $D = 1$, as expected from their difference in geometry.

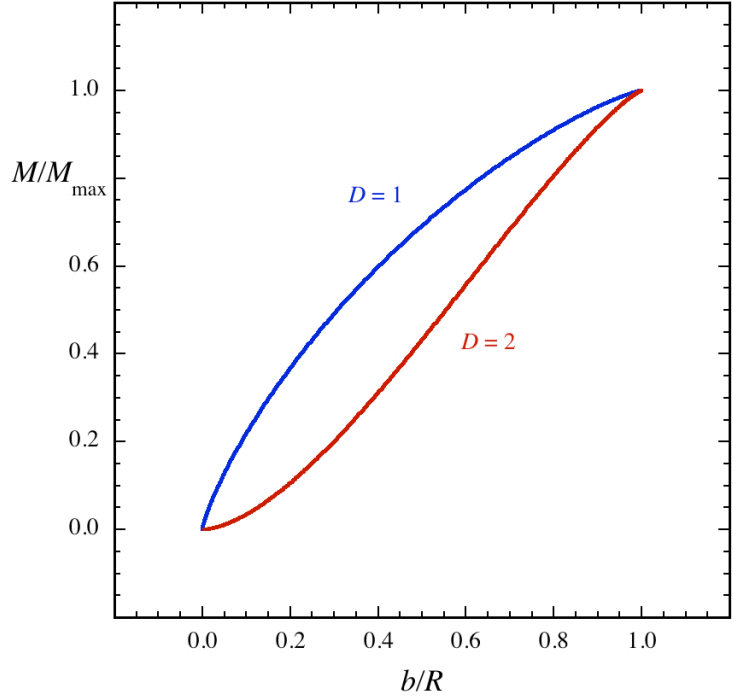

**Figure 3**. Mass enclosed by the projected radius $b$, for condensations characteristic of a lognormal $N$-pdf having width parameter $\sigma_\eta = 0.5$ as in Figure 1, whose projected structure has circular symmetry ($D = 2$) or linear symmetry ($D = 1$).

The normalized half-mass radius $\beta_{1/2}$, which encloses half the condensation mass, is obtained by setting $\lambda_{LN}(\beta) = 1/2$ in equation (8), giving

$$\beta_{1/2} = \left[\frac{1 - \mathrm{erf}(\sigma_\eta/\sqrt{2})}{2}\right]^{1/D} . \tag{9}$$

Equation (9) indicates that the characteristic filament becomes more centrally condensed as $\sigma_\eta$ increases, with $\beta_{1/2}$ decreasing from 0.5 to 0.2 as $\sigma_\eta$ increases from 0 to 1. This behavior is expected, since increasing $\sigma_\eta$ puts more mass at high and at low column density. For typical observed values of $\sigma_\eta = 0.4 - 0.6$, the half-mass radius is $\beta_{1/2} = 0.3$ for $D = 1$ and $\beta_{1/2} = 0.5\text{-}0.6$ for $D = 2$. These values indicate relatively modest central concentration compared to the power-law cases considered next.

2.3  Inversion of a lognormal $N$-pdf with a power-law tail

In many star-forming clouds, the $N$-pdf departs from normal shape, and instead has a power-law tail at high column density. The tendency toward this LNPL shape is seen in regions of low-mass star formation and in more massive regions with embedded clusters (Kainulainen et al. 2009, Schneider et al. 2014b). This property is also seen in simulations of turbulent fragmentation as the gas density increases (Vazquez-Semadeni 1994, Padoan et al. 1997, Federrath & Klessen 2012), and as time proceeds (Ward et al. 2014).

The characteristic LNPL profile $N_{pdf}(\beta)$ is derived in the same way as in Sections 2.1 and 2.2 above. The $N$-pdf follows a normal distribution with mean $\eta_0$ and standard deviation $\sigma_\eta$, for values of $\eta$ in the range $-\infty \leq \eta \leq \eta_1$, and it follows a declining power law with slope $-\Gamma$ for $\eta_1 \leq \eta \leq \infty$. It is assumed that the transition point $\eta_1$ is at least as great as the mean $\eta_0$ of the normal distribution so that the model $N$-pdf resembles observed $N$-pdfs. It is also assumed that $\Gamma > 1$ so that the enclosed mass of the characteristic filament increases outward.

Regions whose $N$-pdfs have significant power-law tails tend to also have prominent filamentary structure (Kainulainen et al. 2009, André et al. 2010). Many filaments host dense cores with smaller aspect ratio, but these cores occupy a relatively small fraction of the filament area. Therefore in this section the inversion is carried out only for structures with linear symmetry, i.e. with $D = 1$.

2.3.1  LNPL $N$-pdf model

The LNPL $N$-pdf is written

$$p(\eta) = p_{LN}(\eta) = A\exp(-u^2), \qquad -\infty \le \eta \le \eta_1 \qquad (10)$$

and

$$p(\eta) = p_{PL}(\eta) = A\exp(-u_1^2)\exp[-\Gamma(\eta - \eta_1)], \qquad \eta_1 \le \eta \le \infty \qquad (11)$$

where $u$ is defined in equation (6) and where $u_1$ is the value of $u$ at the LN - PL transition point. The normalization constant $A$ is determined by requiring $\int_{-\infty}^{\infty} d\eta\, p(\eta) = 1$, giving

$$A = \left\{ \left(\frac{\pi}{2}\right)^{1/2} \sigma_\eta [1 + \text{erf}(u_1)] + \left(\frac{1}{\Gamma}\right)\exp(-u_1^2) \right\}^{-1}. \qquad (12)$$

When the power-law component becomes negligible, $u_1 \to \infty$ and equation (12) reduces to $A = (\sqrt{2\pi}\sigma_\eta)^{-1}$ as expected for the pure normal distribution.

The value of $\eta_0$ needed to obtain $u$ and $u_1$ is determined as in Section 2.2 by requiring that the mean of $N/\overline{N}$ over the function $p(N/\overline{N})$ equal unity. This condition leads to the relation

$$1 = A\left\{ \left(\frac{\pi}{2}\right)^{1/2} \sigma_\eta \exp(\eta_1 - \sqrt{2}\sigma u_1 + \sigma^2/2)[1 + \text{erf}(u_1 - \sigma/\sqrt{2})] + \frac{\exp(\eta_1 - u_1^2)}{\Gamma - 1} \right\} \qquad (13)$$

where the first and second term on the right-hand side represent the fraction of the area whose column densities respectively follow a LN or a PL function. Once equation (13) is solved for $u_1$, $\eta_0$ is obtained from $\eta_0 = \eta_1 - \sqrt{2}\sigma_\eta u_1$ as in equation (6). Alternatively, $\eta_0$ can be estimated from the mode of the $N$-pdf, derived from the normal-function fit to the $N$-pdf for $-\infty \le \eta \le \eta_1$.

2.3.2. LNPL column density profile

The log of the column density profile, $\eta(\beta) = \ln[N(\beta)/\overline{N}]$, is obtained by integrating $p(\eta)$ in equation (3), using equations (10) and (11). For integration limits $\eta(l) = -\infty$, $\eta(u) = \eta$, $\beta(l) = \beta$, and $\beta(u) = 1$, the LN component is

$$\eta_{LN}(\beta) = \eta_0 + \sqrt{2}\sigma_\eta \text{erf}^{-1}\left[\frac{1-\beta}{\sqrt{\pi/2}A\sigma_\eta} - 1\right], \qquad -\infty \le \eta \le \eta_1, \ \beta_1 \le \beta \le 1. \qquad (14)$$

For limits $\eta(l) = \eta$, $\eta(u) = \infty$, $\beta(l) = 0$, and $\beta(u) = \beta$, the PL component is

$$\eta_{PL}(\beta) = \eta_1 - \left(\frac{1}{\Gamma}\right)\ln\frac{\beta}{\beta_1}, \qquad \eta_1 \le \eta \le \infty, \ 0 \le \beta \le \beta_1 \qquad (15)$$

where $\beta_1$ is the coordinate corresponding to the transition point $\eta_1$. The value of $\beta_1$ is obtained by requiring $\eta(\beta_1) = \eta_1$ in equation (15), in combination with equation (12), giving

$$\beta_1 = \frac{A\exp(-u_1^2)}{\Gamma}. \qquad (16)$$

The column density profile $N_{pdf}(\beta)/\overline{N} = \exp[\eta(\beta)]$ is obtained from equations (15) and (16). It declines as a power law for $0 \leq \beta \leq \beta_1$ and then truncates to zero in the range $\beta_1 \leq \beta \leq 1$. When the power-law component becomes negligible as $u_1 \rightarrow \infty$, the profile in equation (16) approaches the pure LN component in equation (7), as expected.

Equation (16) shows that the profile $N_{pdf}(\beta)$ derived from a LNPL $N$-pdf having PL slope $-\Gamma$ also has a PL component, with slope $-1/\Gamma$. This component is an exact power law, in contrast to the approximate power law behavior of the LN $N_{pdf}(\beta)$. A similar relation has been noted between the power-law slope of the $N$-pdf and the exponent of the corresponding volume density profile (Federrath & Klessen 2013).

Equation (16) also shows that the power-law portion of the column density profile typically extends for a small portion of the maximum projected radius. For typical values $A = 1$, $u_1 = 1.3$, and $\Gamma = 2$, $\beta_1 = 0.09$. Nonetheless, this innermost region of the characteristic structure can contain a significant fraction of its mass.

Examples of a LNPL $N$-pdf and its corresponding column density profile $N_{pdf}(\beta)$ are given in **Figures 4** and **5**. **Figure 4** shows a LNPL $N$-pdf having the same width as the LN $N$-pdf in Figure 1, with a power law of slope -1.31 as in NGC3603 (Schneider et al. 2014a), which joins near the peak of the $N$-pdf as in Orion B (Lombardi et al. 2014). This $N$-pdf is chosen because its shallow slope and joining point near its peak yield a characteristic filament with high central concentration. **Figure 5** shows the derived LNPL $N_{pdf}(\beta)$, with the corresponding pure LN $N_{pdf}(\beta)$ and the corresponding pure PL $N_{pdf}(\beta)$ for comparison.

In Figure 5, the PL and LNPL profiles are more centrally concentrated than the LN profile. The PL and LNPL profiles are nearly identical at small radius, but the PL profile is slightly greater at large radius because it does not roll off to zero as does the LN component of the LNPL profile.

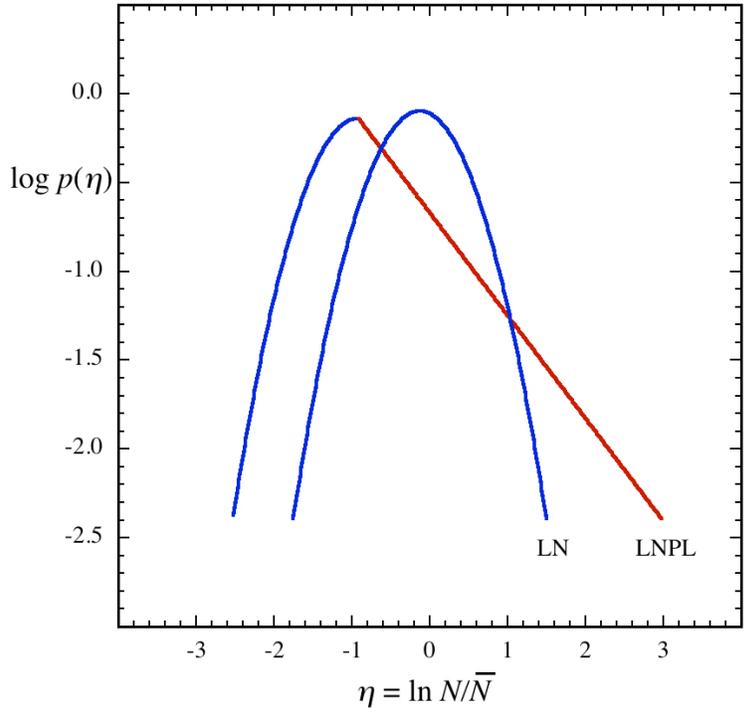

**Figure 4.** LNPL and LN *N*-pdfs having width parameter $\sigma_\eta = 0.5$ as in Figure 1. The LNPL power law joins the *N*-pdf at its peak and has log-log slope -1.3.

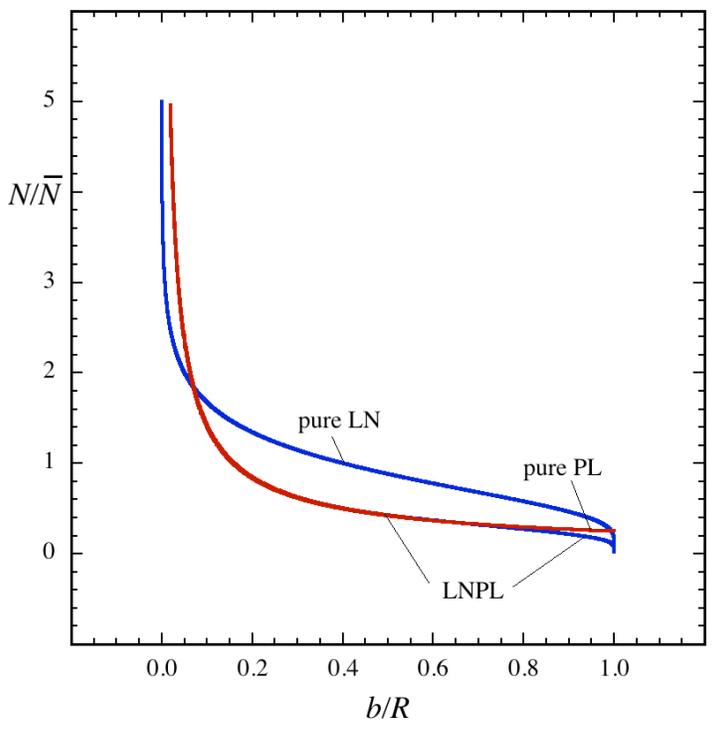

**Figure 5.** Column density *N* as a function of projected radius *b*, for filaments characteristic of the LN and LNPL *N*-pdfs in Figure 4, and for a pure PL *N*-pdf having the same slope as the PL portion of the LNPL *N*-pdf. Red curves indicate PL gas, blue curves indicate LN gas. The column density is normalized to its mean value, and the projected radius is normalized to its maximum value.

2.3.3. LNPL enclosed mass profile

The enclosed mass of the PL component is obtained by integrating the column density in equation (15) over the coordinate extent as in Section 2.2.3, giving

$$\lambda_{PL}(\beta) = \frac{\exp(\eta_1)\beta_1^{1/\Gamma}\beta^{1-1/\Gamma}}{\Gamma - 1} \qquad (17)$$

for $0 \leq \beta \leq \beta_1$. Note that as $\beta \to \beta_1$, equation (17) gives the mass fraction of all of the power-law gas,

$$\lambda_{PL}(\beta_1) = \frac{\exp(\eta_1)\beta_1}{\Gamma - 1}. \qquad (18)$$

Similarly the enclosed mass of the LN component for $\beta_1 \leq \beta \leq 1$ is obtained by integrating the column density in equation (15), giving

$$\lambda_{LN}(\beta) = \left(\frac{\pi}{2}\right)^{1/2} A\sigma_\eta \exp\left(\eta_0 + \sigma_\eta^2/2\right)\left[E(\beta) - E(\beta_1)\right] \qquad (19)$$

where

$$E(\beta) \equiv \text{erf}\left[\frac{\sigma_\eta}{\sqrt{2}} - \text{erf}^{-1}\left(\frac{1-\beta}{\sqrt{\pi/2}A\sigma_\eta} - 1\right)\right] \qquad (20)$$

and

$$E(\beta_1) = \text{erf}\left[\frac{\sigma_\eta}{\sqrt{2}} - u_1\right] . \qquad (21)$$

The enclosed mass of all the LN gas follows from equations (19-21) when $\beta = 1$, giving

$$\lambda_{LN}(1) = \left(\frac{\pi}{2}\right)^{1/2} A\sigma_\eta \exp(\eta_0 + \sigma_\eta^2/2)\left[1 + \text{erf}(u_1 - \sigma_\eta/\sqrt{2})\right]. \qquad (22)$$

The expressions for the enclosed PL and LN mass in equations (18) and (22) are essentially the same as the second and first terms of the cumulative probability of LN and PL column density in equation (13), after a simple algebraic substitution. This agreement verifies the consistency of the inversion equations, since equation (13) is derived by integrating over log column density $\eta$, while equations (18) and (22) are derived by integrating over coordinate $b$.

As expected from the column density profiles, the LNPL mass profile has more of its mass at small radius than the LN mass profile, or than a uniform profile. **Figure 6** shows enclosed mass profiles for the filaments characteristic of the LN, LNPL, and PL $N$-pdfs discussed above. For comparison the mass profile of a uniform column density filament is also shown. Figure 6 shows that at $\beta = 0.01$, the mass enclosed by the LNPL or PL filament is 0.3 of its total mass, and this enclosed mass exceeds that of the uniform filament by a factor of ~30. In contrast the mass enclosed by the LN filament is 0.03 of its total mass, and this mass exceeds that of the uniform column density filament by a factor ~3.

As in Section 2.2 for the LN case, a simple expression for the half-mass LNPL radius exists, provided the power-law component of the mass fraction profile exceeds 1/2, as in Figure 6. Setting equation (17) to 1/2 gives

$$\beta_{1/2} = \left[2Q \exp(\eta_1)\beta_1^{1/\Gamma}\right]^{-Q} \tag{23}$$

where

$$Q \equiv \frac{\Gamma}{\Gamma - 1} . \tag{24}$$

The analytic expressions in equations (9) and (23) for the half-mass radius are useful, since each of them gives a single number which indicates the degree of central concentration of the characteristic filament, without requiring computation of its full structure. Applying equations (9) and (23) to the LNPL, LN, and uniform filaments in Figure 6 gives respective half-mass radii $\beta_{1/2} = 0.053, 0.31,$ and 0.50. These values verify the curves in Figure 6, and they quantify the relative mass concentrations in the LNPL, LN and uniform filaments.

When the enclosed mass fraction of power-law gas in equation (18) is less than 1/2, equation (23) does not apply. Instead, $\beta_{1/2}$ can be obtained by setting the sum of equation (18) and (19) to 1/2, with $\beta = \beta_{1/2}$ in equation (19).

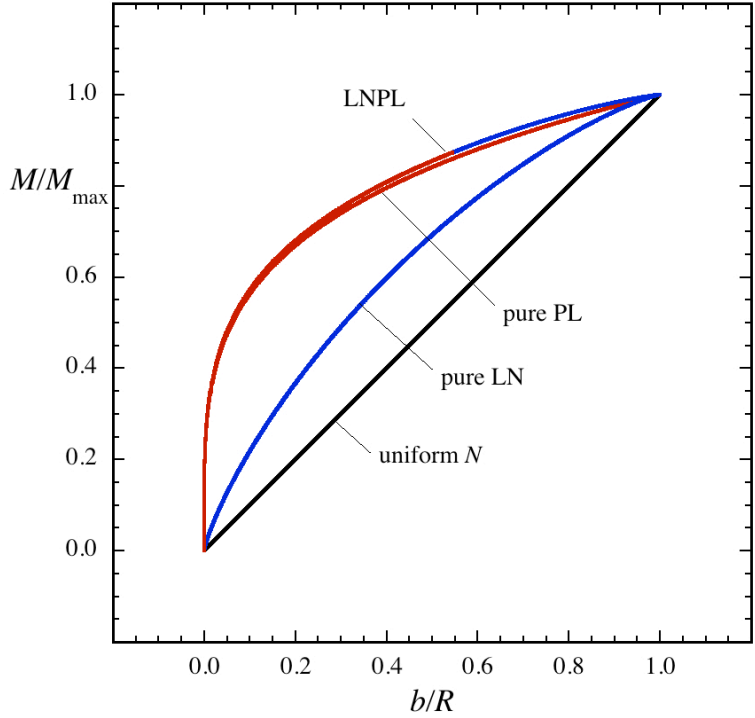

**Figure 6.** Enclosed mass profiles for the filaments characteristic of the LNPL and LN *N*-pdfs in Figures 1 and 3. For comparison the enclosed mass profile of a uniform column density filament is also shown. Blue curves indicate LN gas and red curves indicate PL gas. The LNPL and PL filaments have more enclosed mass at small radius than does the LN filament or the uniform filament.

The half-mass radius $\beta_{1/2}$ depends on the parameters of the LNPL *N*-pdf $\eta_1$, $\sigma_\eta$, and $\Gamma$. It is a more effective indicator of the concentration of a LNPL characteristic filament than is any one of these parameters alone. The slope $\Gamma$ indicates how steeply the power-law gas is concentrated, but the departure of the attachment point from the peak, in units of the LN width, is also needed to determine the fraction of the *N*-pdf gas which is so concentrated. The half-mass radius combines these properties into a single diagnostic number. For example, increasing the attachment point from $u_1 = 0$ to 2 in the LNPL example in Figure 6, with $\sigma_\eta = 0.5$ and $\Gamma = 4/3$, increases the half-mass radius significantly, from 0.059 to 0.25, because then the power-law gas occupies a smaller fraction of the entire *N*-pdf.

2.4. Inversion of a power-law *N*-pdf

If the *N*-pdf is a pure power law (PL) with no significant LN component as in the IRDC observations of Schneider et al. (2014*b*), the same analysis as in the foregoing sections can be applied. The PL inversion is a special case of the LNPL inversion, where the LN parameters $u_1$ and $\sigma_\eta$ each approach zero, whence $A \to \Gamma$ and $\beta_1 \to 1$. If the model *N*-pdf follows a pure power law of slope $-\Gamma$ where $\eta_1 \le \eta \le \infty$, then after normalization as in Section 2.3.1, the *N*-pdf becomes

$$p(\eta) = \Gamma \exp[-\Gamma(\eta - \eta_1)] \qquad (23)$$

and the log column density profile becomes

$$\eta = \eta_1 - \frac{1}{\Gamma} \ln \beta \qquad (24)$$

for $0 \le \beta \le 1$, where

$$\eta_1 = \ln \frac{\Gamma - 1}{\Gamma}. \qquad (25)$$

Then the column density profile is simply

$$\frac{N}{\bar{N}} = \frac{\Gamma - 1}{\Gamma} \beta^{-1/\Gamma}, \qquad (26)$$

the enclosed mass profile becomes

$$\lambda_{PL}(\beta) = \beta^{1-1/\Gamma} \quad , \tag{27}$$

and the half-mass radius becomes

$$\beta_{1/2} = 2^{-1/(1-1/\Gamma)} \quad . \tag{28}$$

Figures 5 and 6 show that when the PL component dominates the LNPL $N$-pdf (when $\beta_1 > \beta_{1/2}$), equations (26)-(28) are good approximations to the properties of the LNPL profiles of column density and mass. Equations (26)-(28) are also simpler and more convenient to use than the corresponding LNPL equations, since they depend only on one parameter, the slope $\Gamma$. For the above example where $\Gamma = 4/3$, equations (26-28) give the column density fraction as $(1/4)\beta^{-3/4}$, the enclosed mass fraction as $\beta^{1/4}$, and the normalized half-mass radius as $2^{-4}$.

2.5. Maximum radius of characteristic filament

A typical value of the maximum filament radius $R$ can be estimated by assuming that the characteristic filament is a spatial average over a self-gravitating bundle of "fibers" (Hacar & Tafalla 2011), as discussed in Section 2.1. *Herschel* observations of B211-B213 in Taurus appear consistent with fibers observed with finer resolution (Palmeirim et al. 2013).

The radius $R$ can be evaluated from the virial radius of the outermost fiber which is bound to the enclosed bundle mass. If the virial speed $\sigma_V$ of this fiber exceeds the isothermal sound speed $\sigma$ by the Mach factor $\mu$, the enclosed mass per length can be written $\Lambda = 2mR\overline{N} = \mu^2 \Lambda_c$ where $m$ is the mean molecular mass and where $\Lambda_c = 2\sigma^2/G$ is the critical mass per length for an infinite isothermal cylinder (Ostriker 1964). Then solving for $R$ yields

$$R = \frac{\sigma_V^2}{mG\overline{N}} \quad . \tag{29}$$

The virial speed of the typical outermost fiber is estimated as 0.5 km s$^{-1}$, based on the largest excursions in line-of-sight velocity from the local mean in the Taurus filaments 15 and 20 (Hacar et al. 2013). For 10 K gas, the Mach number of this speed is $\mu = 2.6$. For a typical value of mean column density in nearby cloud $N$-pdfs, $\overline{N} = 3 \times 10^{21}$ cm$^{-2}$, equation (29) gives the maximum radius as $R = 1.0$ pc. A similar value $R = 1.5$ pc was adopted for the radial extent of the well-studied filament #6 in IC5146 (Arzoumanian et al. 2011).

It may also be possible to estimate a representative value of $R$ from a column density map of a filamentary cloud, by decomposing the map contours into prolate figures as in Kainulainen et al. (2014). Then the distribution of map area as a function of $R$ can be used to find the modal value of $R$, i.e. the value of $R$ whose associated prolate figures occupy the greatest fraction of the map area.

2.6. Mean density within the half-mass radius

The estimated typical maximum radius allows an estimate of the mean density within the half-mass radius, for a cylindrical filament. This estimate assumes that the half-mass cylindrical radius $r_{1/2}$ equals the half-mass projected radius $b_{1/2}$, as is strictly true for $r_{1/2} \ll R$ and $b_{1/2} \ll R$. The mean density is then

$$\overline{n}_{1/2} = \frac{\overline{N}}{\pi \beta_{1/2}^2 R} \quad . \tag{30}$$

For $\overline{N} = 3 \times 10^{21}$ cm$^{-2}$, $R = 1.0$ pc, and for the LN and LNPL half-maximum radii in section 2.3.3 above, the mean density within the half-mass (HM) radius is $3 \times 10^3$ cm$^{-2}$ for the LN filament and $1 \times 10^5$ cm$^{-2}$ for the LNPL filament discussed above. The typical mean density of star-forming dense cores in nearby clouds is of order a few $10^4$ cm$^{-3}$ based on observations of the dust continuum (Enoch et al. 2007, Evans et al. 2009), and of NH$_3$ lines (Myers & Benson 1983). For the LN and LNPL cases considered above, the gas within the HM filament radius is much denser and thus more likely to harbor star-forming cores than is the gas within the HM radius of the LN filament.

2.7. Star-forming fertility

The "fertility" of gas for star formation indicates that a filamentary fiber contains one or more regions of emission in the 1-0 line of $N_2H^+$, a tracer of cores dense enough to form one or more low-mass stars (Hacar & Tafalla 2011). This concept is useful to distinguish "fertile" fibers, which have $N_2H^+$ emission, from "sterile" fibers, which do not have $N_2H^+$ emission. Fertility differs from star formation efficiency (SFE) in that fertility refers to the potential of a region to form stars in the near future, rather than the efficiency with which a region has formed stars in the recent past. Thus a region which is just beginning to form stars may have low SFE but high fertility, and a region which has used up its supply of dense gas in forming stars may have high SFE but low fertility. Embedded clusters, which have many young stars and are still forming new stars, should have both high SFE and high fertility.

Here the basic idea of fertility is given a quantitative form for a characteristic filament. It applies only to filaments which are dense and unstable enough to collapse, and it distinguishes them by their mean mass accretion rates during collapse. If the gas within the half-mass radius of a cylinder collapses in free fall from rest, without fragmenting, its mean mass accretion rate per unit length $\dot{\Lambda}_{1/2}$ is estimated in terms of its total mass per unit length $\Lambda$ as

$$\dot{\Lambda}_{1/2} = (\Lambda/2)\left(4Gm\bar{n}_{1/2}\right)^{1/2} \qquad (31)$$

using the expression for the free fall time of a collapsing cylinder (Penston 1969). Combining equations (30-31) gives the mean mass accretion rate per length of the collapsing gas initially within the half-mass radius of a filament,

$$\dot{\Lambda}_{1/2} = 2mN_1 \left(\frac{G\Lambda}{\pi}\right)^{1/2} f \ . \tag{32}$$

Here the fertility $f$ is the normalized mean mass accretion rate per length due to free fall collapse of gas within the half-mass radius of a condensed filament. This rate is normalized by the mean mass accretion rate per length of the gas initially within the half-mass radius of a filament having the same mass per length, but which has minimal column density and concentration. It is initially uniform, and its mean column density $N_1 = 1 \times 10^{21}$ cm$^{-2}$ approximates the lowest value which can reliably be measured. Then

$$f \equiv \frac{\overline{N}/N_1}{2\beta_{1/2}} \ . \tag{33}$$

Equations (32-33) show that a filament of given mass per length has the greatest fertility if it has both high mean column density and high central concentration. The mass accretion rate of collapsing gas within the half-mass radius of a filament of mass per length 20 $M_\odot$ pc$^{-1}$, mean column density $\overline{N} = N_1 = 1 \times 10^{21}$ cm$^{-2}$ for a segment of length 1 pc is $6.28 f \ M_\odot$ Myr$^{-1}$.

For a pure PL $N$-pdf the fertility of the corresponding characteristic filament is $f = 2^{1/(\Gamma-1)} \overline{N}/N_1$ from equations (28) and (33). Values of $f$ extend from $f = 1$ for $\overline{N} = N_1$ and $\Gamma \to \infty$, to unbounded values as $\Gamma \to 1$. Among clouds with well-measured $N$-pdfs, $f = 28$ for the shallowest reported value of $\Gamma = 1.31$ in NGC 3603, where $\overline{N}/N_1 = 3$ (Schneider et al. 2014a). The shallowest reported value of $\Gamma$ does not necessarily give the greatest fertility, however, since a greater value $f = 53$ is seen for the greatest reported value of $\overline{N}/N_1 = 39.9$ in the IRDC G28.37, where $\Gamma = 3.42$ (Schneider et al. 2014b).

## 3. Characteristic filament properties

This section compares the characteristic filament structure for LNPL and LN pdfs with cylindrical filament models.

3.1. LN filaments

For linear symmetry with $D = 1$, which approximates the elongation of many observed filaments, the LN $N_{pdf}(\beta)$ can be approximately fit by the $N$-profile of a cylinder of finite radius $R$, whose volume density $n$ declines with cylindrical radius $r$ over $0 \leq r \leq R$ as a power law, $n \sim r^{-p}$ ($p > 0$). For typical observed values of $\sigma_\eta = 0.4 - 0.6$, $N_{pdf}(\beta)$ corresponds to a relatively shallow power-law decline, $p = 0.6 - 1.0$. **Figure 7** shows a close match between $\eta(\beta)$ for $\sigma_\eta = 0.5$ and for a power-law cylinder with $p = 0.8$. The power-law cylinder matches because at mid-range values of $\beta$, the cylinder column density declines approximately as a power law of $\beta$, and also because as $b$ approaches the cylindrical boundary radius, the cylinder geometry truncates the column density to zero.

Although the variation $N_{pdf}(\beta)$ for a LN $N$-pdf is not an exact power law, its departure is relatively small (Section 2.2.2). On the other hand, one cannot simply estimate the density power law exponent $p$ from the column density log-log slope by adding unity, as can be done in the limit of small radius when the column density declines with projected radius as a true power law. Instead it is necessary to assume a density power law exponent, integrate to obtain the corresponding column density profile, and compare with the characteristic profile.

These slightly condensed power-law cylinders are more condensed than a uniform density cylinder, as also shown in Figure 7. The geometry of a uniform density cylinder gives a slight decrease of log column density with increasing log radius, and then a sharp decline as $b$ approaches $R$.

The case $p = 1.0$ when $\sigma_\eta = 0.6$ corresponds to a "logatropic" cylinder with polytropic exponent $\gamma_p \rightarrow 0$ (Gehman et al. 1996, TG15). For more common values of $\sigma_\eta$ in the range 0.3 - 0.5, the fits have $p < 1$, which implies $\gamma_p < 0$. They do not correspond to any cylindrical polytrope, since for $\gamma_p < 0$ the polytropic pressure $P \propto \rho^{\gamma_p}$ decreases with increasing density. Thus most regions whose $N$-pdf is LN have characteristic filaments which are not centrally condensed enough to be self-gravitating.

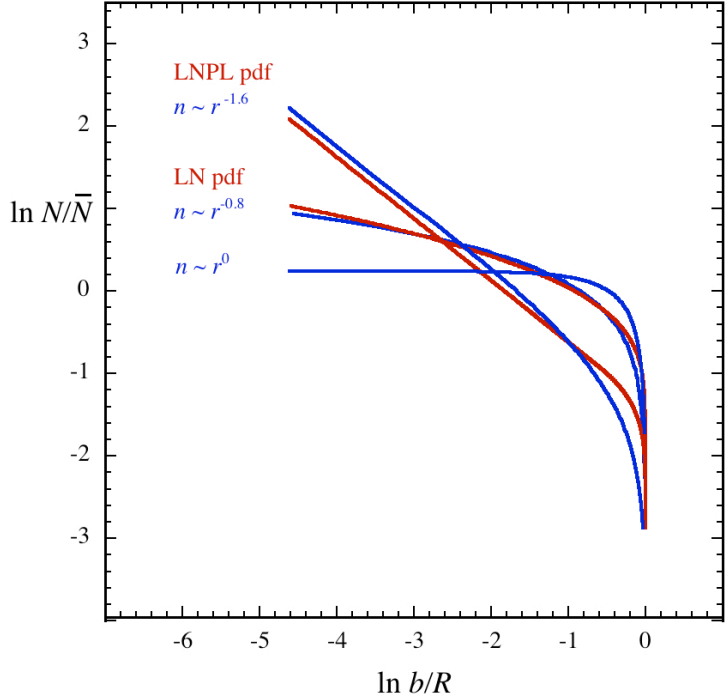

**Figure 7.** Column density profiles for characteristic LNPL and LN filaments described in Figures 5 and 6 (*red*), and for power-law cylinder models having volume density $n$ decreasing with cylindrical radius $r$ as $n \sim r^{-p}$, with $p = 1.6, 0.8,$ and $0$ (*blue*).

3.2. LNPL filaments

FIlaments characteristic of a $N$-pdf having LNPL shape also have a column density profile $N_{\rm pdf}(\beta)$ approximating that of a truncated power law of volume density, but with a steeper density exponent than for the pure LN pdf, as expected from the comparisons in Section 2. For the same LNPL pdf as in Figures 5 and 6, Figure 7 shows that the matching cylinder profile has density exponent $p = 1.6$, as compared with $p = 0.8$ for the LN pdf. Unlike the LN case, the roll-off of the cylinder profile at large radius is too gradual to fully match the LNPL column density profile. This gradual rolloff appears due to the assumed cylinder geometry and not to the density power law. A better match can be achieved by assuming that the cylinder cross-section is a flattened circle, i.e. more ellipsoidal than round.

This density exponent $p = 1.6$ corresponds to polytropic index $n = -4$ or polytropic exponent $\gamma_p = 2-2/p = 0.75$ (TG15). Since this LNPL $N$-pdf has a shallow power-law tail with attachment point near the peak, its characteristic filament is more centrally condensed than for most $N$-pdfs. Thus profiles arising from inversion of most LNPL pdfs will have cylinder density exponents smaller than 1.6, i.e. $p = 1.0 - 1.6$. They will therefore match self-gravitating models with polytropic exponents $\gamma_p = 0 - 0.75$.

As the LNPL $N$-pdf approaches a pure power law, the polytropic exponent of the corresponding singular cylinder can be estimated more simply than in Figure 7. When the change in slope at large radius due to truncation can be neglected, the exponent $p$ in the volume density power law $n \sim r^{-p}$ exceeds the exponent $q$ in the column density power law $N \sim b^{-q}$ by unity. Then the polytropic exponent $\gamma_p$ is directly related to the exponent $\Gamma$ in the PL component of the $N$-pdf by $\gamma_p = 2/(1+\Gamma)$, using the relation between $p$ and $\gamma_p$ for singular cylindrical polytropes (TG15). The PL slopes $\Gamma$ of observed $N$-pdfs generally exceed unity, further indicating that singular polytropic models of characteristic filaments are sub-isothermal, i.e. they have $\gamma_p < 1$.

A similar range of sub-isothermal polytropic exponents was inferred for regular cylindrical polytropes, by considering observed filament column density profiles (TG15). Both singular and regular polytropic models describe a self-gravitating equilibrium configuration. However for these two models, only the singular polytrope is consistent with the characteristic filament of a $N$-pdf having PL or LNPL shape. Application of equation (1) to a regular polytrope such as the Plummer cylinder model with $p = 2$ (Arzoumanian et al. 2011) yields a $N$-pdf which clearly departs from observed $N$-pdfs, by increasing as the column density approaches its maximum value (Fischera 2014). This departure is due to the nearly constant density within the first thermal scale length, a property of all regular polytropes.

In the limit where the $N$-pdf is a pure PL, for the above example the column density profile is $N_{pdf}(\beta) = (1/4)\beta^{-3/4}$ as given in Section 2.4. This PL profile is essentially the same as the LNPL profile in Figure 7, except that its curve remains straight and does not roll off as $\beta \rightarrow 1$. Thus it is also well fit by the cylinder power law with volume density $n \sim r^{-1.6}$. The exponent 1.6 is slightly less than $1+1/\Gamma = 1.75$ expected for a pure power law, due to the rolloff imposed by the cylindrical geometry.

If most filaments have more complex internal structure suggested by the fibers in Taurus and other nearby clouds (Hacar et al. 2011), the similarity to simple cylindrical polytropes as discussed by Arzoumanian et al. (2011) and TG15 may be useful mainly as a large-scale average. If so, it may not be possible to distinguish from current observations alone whether singular or regular polytrope models are better descriptions of internal filament structure. Spectral-line observations in lines which probe a range of density should be more revealing of the dynamical status of filaments.

## 4. Application to observed $N$-pdfs

This section illustrates the derivation of characteristic filament structures from observed $N$-pdfs. These structures differ from the characteristic filaments derived in Section 3 because they are more realistic, and because the observational uncertainties in their $N$-pdf parameters are used to estimate uncertainties in the inferred normalized half-mass radii. The results indicate that there are measurable differences from region to region in the concentration of the characteristic filament, that these differences lie outside measurement error, and that the corresponding differences in fertility are similar to those in star formation efficiency.

However these results are not based on analysis of a statistically large sample. Once larger samples of carefully selected $N$-pdfs are available, one may extend the calculations presented here, to differentiate star-forming regions and their fertility for future star formation.

### 4.1. Region selection

Characteristic structures are derived for three $N$-pdfs based on *Herschel* observations which are analyzed with the same technique, including careful removal of background emission. They vary in the prominence of their power-law tail, from G216-2.5 with a steep power law attached far from the peak, to the Auriga-California cloud with a shallower power law attached closer to the peak, to NGC3603 with a still shallower power law attached still closer to the peak (Schneider et al. 2014*a*).

Their star formation activity is similarly low for G216-2.5 and Auriga-California, and much higher for NGC3603.

G216-2.5 (Maddalena & Thaddeus 1985) has relatively little star formation for its mass, which exceeds $10^5\ M_\odot$. It has an estimated 33 protostars and 41 young stars with disks, according to infrared observations by the *Spitzer* space telescope and by ground-based telescopes (Megeath et al. 2009). The densest gas and protostars are concentrated in an elongated structure with projected length $\sim 10$ pc. The region mapped by *Herschel* used to construct the *N*-pdf is more extended by a factor $\sim 10$ in each dimension.

The Auriga-California cloud has similar mass to G216-2.5 and harbors 175 young stellar objects (YSOs) according to infrared observations with the *Herschel*, *Spitzer*, and *WISE* telescopes (Lada et al. 2009, Harvey et al. 2013). The cloud gas and dust emission are concentrated in a filamentary network.

NGC3603 is an H II region associated with a massive starburst cluster, hosting more than 7000 young stars (Harayama et al. 2008). It is considered one of four super-star-clusters in the Milky Way still associated with parent dense gas. It is surrounded by two elongated molecular clouds extended over $\sim 20$ pc (Nürnberger et al. 2002), and is near in projection to two additional massive elongated clouds associated with the clusters NGC3576 and NGC3590 (Fukui et al. 2014). All four of these clouds are included in the *N*-pdf reported by Schneider et al. (2014).

4.2. Characteristic filament mass profiles

The equations for mass profile of a characteristic filament in Section 3 were used to compare the characteristic filaments of these three regions, shown in **Figure 8**. The background-corrected parameters used to compute the curves are $(\eta_1, \sigma_\eta, \Gamma) = (0.30, 0.52, 1.31)$ for NGC3603, $(0.65, 0.45, 2.59)$ for Auriga, and $(0.87, 0.32, 3.65)$ for G216-2.5, as given in Table 1 and Figures 1 and 2 of Schneider et al. (2014*a*). The one-sigma uncertainties on $\sigma_\eta$ and $\Gamma$ due to least-squares fitting are (0.02, 0.02) for NGC3603, (0.02, 0.02) for Auriga, and (0.07, 0.02) for G216-2.5 (Schneider 2015, personal communication). These fitting uncertainties are relatively small because the fit functions are good models of the *N*-pdf shape and because each *N*-pdf typically contains thousands of independent pixels. The uncertainties on $\eta_1$ are negligible. Systematic uncertainties due to foreground and background contamination are discussed by Schneider et al. (2014) and Lombardi et al. (2015).

The uncertainty in $\beta_{1/2}$ for G216-2.5 and Auriga was estimated from equation (9), by assuming that the uncertainty is dominated by $\sigma_\eta$, and by assuming that $\sigma_\eta$ departs from its published value by ± its one-sigma uncertainty. Similarly, the uncertainty in $\beta_{1/2}$ for NGC3603 was obtained from equation (23), by assuming that the uncertainty is dominated by $\Gamma$, and by assuming that $\Gamma$ departs from its published value by ± its one-sigma uncertainty.

The mass profiles in Figure 8 show modest central concentration for G216-2.5 and Auriga, and significantly greater concentration for NGC 3603. This trend is evident as the LNPL half-mass radius $\beta_{1/2}$ decreases from $0.37 \pm 0.01$ for G216-2.5 to $0.30 \pm 0.01$ for Auriga to $0.09 \pm 0.01$ for NGC 3603. The central concentration also increases as the power-law component of the profile becomes more important. For G216-2.5, the PL component of the $N$-pdf attaches far from its peak. Thus the PL component plays no significant role and the LNPL half-mass radius is matched exactly by its pure LN value, in equation (9). In contrast, NGC 3603 is dominated by its PL component, since the PL of its $N$-pdf has shallow slope and attaches close to its peak. Thus its LNPL half-mass radius is much more closely matched by the pure PL value, in equation (28), than by its LN value.

This progression of mass profile concentrations allows additional comparisons of the three regions. The pure PL filament having the same half-mass radius matches singular cylinder models with $p = 1.3$, 1.4, and 1.7 for G216-2.5, Auriga, and NGC3603, respectively. These exponent values are based on equation (28) and on assuming that $p = 1 + \Gamma$, as for cylinder radii much smaller than the bounding radius. Singular cylinders with these exponents can be self-gravitating singular cylinder polytropes, ranging from weakly to strongly bound (TG15). Since their exponents lie in the range of bound cylinders, their fertility for future star formation can be estimated from equation (37). Their mean column densities are $\overline{N}/N_1 = 2.1$, 1.5, and 3.2 respectively (Schneider et al. 2014a). Then the fertility is low and comparable between G216-2.5 and Auriga ($f = 2.8, 2.5$), but is much greater for NGC3603 ($f = 18$).

It is well known that presence of a PL tail in the $N$-pdf of a region is associated with increased star formation activity (Kainulainen et al. 2019, Schneider et al. 2014a), and this association is often ascribed to gas which is self-gravitating or collapsing (Federrath & Klessen 2013, Girichidis et al.

2013). However the relative importance of the slope and attachment point as indicators of the star formation efficiency, or of the potential for further star formation, remain unclear. The present analysis resolves the ambiguity between the slope and attachment point in setting the concentration of a LNPL profile. It remains to be studied how well the concentration and the mean column density of a profile are correlated with star formation activity, in a statistically large sample of regions which are selected in an unbiased manner.

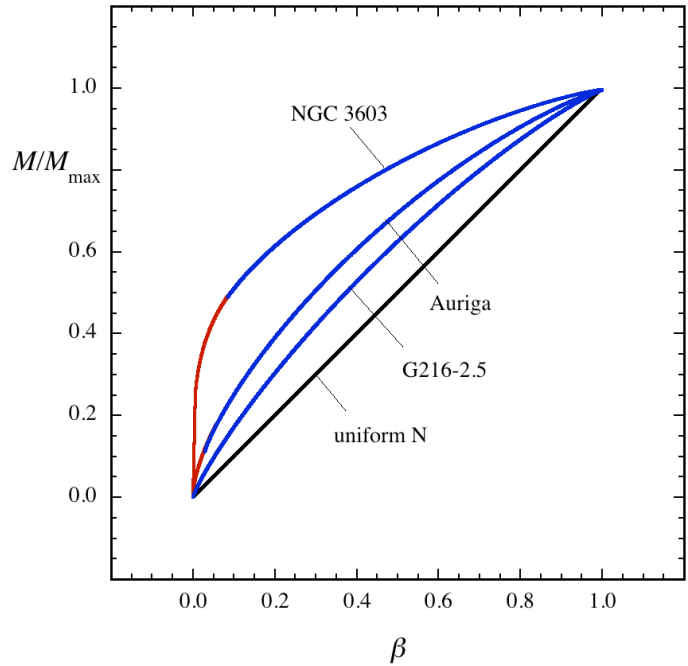

**Figure 8.** Enclosed mass profiles for characteristic filaments of star-forming regions NGC3603, Auriga, and G216-2.5, whose $N$-pdfs are based on *Herschel* observations (Schneider et al. 2014*a*). Each colored curve is based on the LNPL fit by Schneider et al. (2014*a*) to the observed $N$-pdf. The red part of each mass profile arises from the PL component of the $N$-pdf and the blue part arises from the LN component. The black line shows the enclosed mass profile of a filament having uniform column density. The degree of central concentration of each profile increases with the star formation activity of each region.

## 5. Discussion

### 5.1. Limitations

The inversion analysis of Sections 2 and 3 obtains column density profiles associated with simple models of *N*-pdfs observed in molecular cloud maps. The utility of this inversion increases as the projected condensations in the region become more similar in size and shape, and as they occupy an increasing fraction of the map area. At one extreme is a region with a small fraction of its map area in condensations of diverse size and shape. In this case, inverting the *N*-pdf does not add useful information about the structure of the region. At the other extreme is a region dominated by a single condensation, or a region having a large fraction of its map area in condensations of similar size and shape. For these, inverting the *N*-pdf can distinguish the typical condensation of one region from another. It can describe the typical degree of central condensation, as an indicator of the likelihood of further star formation, and its comparison with dynamical models can constrain its dynamical status.

To make best use of the inversion analysis, it may be necessary to select regions whose condensations remain similar and whose *N*-pdfs remain similar, as the region location and area are varied.

Caution is also needed if the condensations in a region have similar structure, but they occupy a small fraction of the region area dominated by lower-density gas. If most of this gas does not lie between the condensations but instead surrounds them, the characteristic filament will have an envelope which is too diffuse and extended. This may occur for LNPL *N*-pdfs in giant molecular clouds (GMCs) having embedded infrared dark clouds (IRDCs), whose IRDCs alone have pure PL *N*-pdfs (Schneider et al. 2014*b*). In this case it may be more useful to evaluate the *N*-pdf above an increased threshold of column density before computing its characteristic structure.

The regions in Figure 8 were selected to demonstrate the methods developed in Sections 2-3, and to illustrate the trend of central concentration with star formation. It remains to be studied how such a trend depends on the way cloud regions are selected, and on the size and composition of the sample.

In comparing *N*-pdfs for their star formation potential it is important to select regions in an unbiased way. For example the *N*-pdf of an IRDC alone will give a characteristic filament which appears more fertile than the characteristic filament of the same IRDC embedded in its surrounding GMC, as in Schneider et al. (2014*b*), mainly because the mean column density of the IRDC exceeds that of the IRDC + GMC, by selection. This bias can be remedied by selecting the same minimum column density for all regions in the sample to be compared.

## 5.2. Dynamical status of characteristic filament

The internal structure of characteristic filaments derived from LN, LNPL, or PL $N$-pdfs in Section 3 is essentially a singular truncated power law of density with cylindrical radius. The power-law slope becomes steeper from LN to LNPL profiles, and its truncation becomes sharper as the PL component becomes more important. For typical parameters of observed $N$-pdfs, LN profile slopes are too shallow to match models of self-gravitating polytropes. This suggests that a region whose $N$-pdf is LN may be infertile for star formation because its typical filament is not concentrated enough to be self-gravitating.

Such an interpretation may apply to regions in the central molecular zone of the Milky Way such as G0.253+0.016 (the "brick"), which has a network of filaments, a nearly LN $N$-pdf, and very little star formation. If the LN width $\sigma_\eta$ is equal to 0.30, as found by Johnston et al. (2014) for SMA and SCUBA data, or 0.34 as found by Rathborne et al. (2014) for ALMA and single-dish data, the characteristic filament has central concentration $\beta_{1/2} = 0.37 - 0.38$. This modest concentration is essentially the same as in the nearly starless GMC G216-2.5 considered above.

In contrast, LNPL and PL profiles are generally steeper, and they match models of self-gravitating singular polytropic cylinders, provided they are cold-centered rather than isothermal, i.e. with polytropic exponent $\gamma_p < 1$. The polytropic exponent can be inferred approximately from $\Gamma$, the magnitude of the power-law slope of the $N$-pdf, by $\gamma_p = 2/(1+\Gamma)$.

As noted in Section 2.2.2, the central structure of the characteristic filament may be uncertain because its underlying $N$-pdf has increased statistical error at high column density. Also, the inversion analysis assumes a monotonic decrease of column density with projected radius, whereas a fine structure of central fibers would have a more complex variation of central column density. Therefore, the conclusion that a characteristic filament matches a polytrope which is singular, rather than regular, requires more investigation. For this purpose it may be useful to evaluate $N$-pdfs based on continuum observations with finer spatial resolution, and on line observations tracing low and high density.

These uncertainties about central density do not apply to the conclusion that the characteristic LNPL or PL filament matches a cold-centered polytropic model, discussed in Section 3.2. This result is based on power-law slopes which are usually well-determined from observational fits to $N$-pdfs. Furthermore, this conclusion is similar to the inference that many observed individual filaments resemble cold-centered polytropes, based on matching their average column density profiles to models of regular polytropes (Arzoumanian et al. 2011, TG15).

It remains to investigate the physical basis of the gradient in polytropic temperature needed to match the typical polytropic exponent ($\gamma_p$ = 0.5-0.7) of the characteristic filament. The typical increase in gas kinetic temperature from center to edge of observed filaments is at most ~ 10 K, too small to match the expected increase in polytropic temperature. Only for polytropic exponent very close to unity ($\gamma_p$ = 0.97) can the polytropic model match observed filament kinetic temperature gradients (Palmeirim et al. 2013). Possible mechanisms of nonthermal support, particularly due to small-amplitude Alfvén waves having $\gamma_p$ = 1/2, are discussed by TG15.

Although the polytropic cylinders considered here are equilibrium bodies, they are singular, so their equilibrium is unstable. Thus they are maximally susceptible to collapse and fragmentation. The corresonding regular polytropes are also unstable against radial collapse for the range of $\gamma_p$ inferred from filament observations (TG15). Thus the self-gravitating models which match LNPL and PL characteristic filaments have a dynamical status which seems appropriate for regions which are on the verge of star-forming collapse.

The static equilibrium interpretation of LNPL and PL characteristic filaments adopted here is incomplete because it does not account for the flows and accretion which are needed to understand how filaments form and grow (Heitsch & Hartmann 2014, Heitsch 2013), and it does not account for filament dispersal due to star formation, photoevaporation, and to winds and outflows. However it offers a physical way to distinguish the star-forming potential of regions having LN $N$-pdfs from those having LNPL and PL $N$-pdfs, and it offers a simple explanation for the longevity of concentrated filaments in star-forming regions.

One might speculate that in a more complete picture, a LNPL or PL filament with typical parameters has the longevity and approximate structure of a cold-centered equilibrium cylinder, even as it is collapsing, fragmenting into cores, and forming stars on the inside, and as it is accreting new mass on the outside. Such characteristic filaments, and their corresponding regions, should differ in their likelihood of star formation according to the fertility indicated by their $N$-pdfs. In contrast a LN filament whose width is in the typical range $\sigma_\eta$ = 0.3-0.5 has too little concentration to be self-gravitating. It should disperse on a sound crossing time scale unless it is confined by external pressure.

It should not be able to collapse by self-gravity to form cores and stars unless it accretes more mass and becomes more centrally condensed.

## 6. Summary and conclusions

This paper investigates the relation between two features of molecular clouds - their probability density functions of column density (*N*-pdfs) and their filamentary structure. These features are related because much of the gas in the denser part of a *N*-pdf is in enlongated filaments. For a given cloud region, the change of variables relation between probability density functions is used to infer the column density profile $N(\beta)$ as a function of normalized projected radius $\beta$, for a single "characteristic" filament having the same *N*-pdf as the region. If the region area is dominated by one filament, or by multiple filaments having similar radial profile, the characteristic filament may distinguish regions, the dynamical status of their dense gas, and their fertility for star formation.

The paper presents analytic expressions for the normalized profile of characteristic filament column density and enclosed mass fraction, for three commonly observed forms of *N*-pdf: a "lognormal" function (LN), a lognormal with a power-law tail at high density (LNPL), and a pure power-law (PL). The degree of central concentration for each profile is quantified by its half-mass projected radius $\beta_{1/2}$. Filaments which are sufficiently concentrated to be self-gravitating are further distinguished by their "fertility" which is proportional to their mass accretion rate during collapse.

The inversion analysis is applied to *N*-pdfs whose parameters exemplify observed regions. Each column density profile is compared with cylindrical power laws of volume density $n \sim r^{-p}$ to find the closest match. Each of these density power laws is then compared with models of singular cylinder polytropes.

The dynamical status of dense *N*-pdf gas is inferred from dynamical models of cylindrical filaments rather than of spherical cores, because significantly more *N*-pdf gas is in filaments than in cores. The models which are considered allow cylinders which are unbound, bound, and collapsing. Considering only collapse models would yield a star formation rate in the next few free fall times much greater than observed in the past few free fall times.

The inversion analysis is applied to LNPL $N$-pdfs from three GMC regions with similar mass, $\sim 10^5$ $M_\odot$, but with different degrees of star formation. They are G216-2.5 with an estimated 74 young stellar objects (YSOs), the Auriga-California cloud with 175 YSOs, and the GMC associated with NGC3603, a super star cluster having more than 7000 YSOs. Their characteristic filaments have progressively smaller half-mass radii, indicating that their central concentration increases with their star formation efficiency.

The main conclusions are:

1. For LN, LNPL, and PL $N$-pdfs, the characteristic filament column density profile decreases from an infinite central value to zero at a finite maximum radius. The mass fraction profile increases from zero to unity with convex shape, indicating greater central concentration than for a filament of uniform column density.

2. These column density and mass profiles are approximately consistent with radially truncated cylinder models whose volume density declines as a power law of cylinder radius. The typical truncation radius is $\sim 1$ pc, based on a virial model and on observations of well-studied filament velocities.

3. For pure LN $N$-pdfs having typical width parameter $\sigma_\eta = 0.3 - 0.5$, the characteristic filament column density declines with radius slowly, approximately matching a cylinder density law $n \sim r^{-p}$, with $p < 1$. Such cylinders cannot match any cylindrical polytrope model because their gas is not concentrated enough to be self-gravitating. Such structures should not collapse or fragment gravitationally to form cores and stars. When $\sigma_\eta \approx 0.6$ the column density profile can be approximated by a marginally self-gravitating logatropic cylinder, where $p = 1$ and where the polytropic exponent is $\gamma_p = 0$.

4. For pure PL $N$-pdfs with typical negative slope $\Gamma > 1$, the characteristic column density profile is a singular power law with negative slope $1/\Gamma$ and the volume density profile has $p \approx 1 + 1/\Gamma$ at small radius. Thus the characteristic filament of any pure PL $N$-pdf matches a self-gravitating singular cylinder model with $1 < p < 2$. As $\Gamma$ approaches 1 the cylinder becomes more centrally condensed, $p$ approaches 2 and the polytropic exponent $\gamma_p = 2/(1+\Gamma)$ approaches the isothermal value of 1. All such pure PL characteristic filaments can be self-gravitating, with varying central concentration depending on $\Gamma$.

5. Many characteristic filament $N$- profiles match singular polytropic cylinder models which are centrally infinite. These differ from many observed filament $N$-profiles, which approach a finite maximum central density (Arzoumanian et al. 2011) and which match regular polytropic cylinder models (TG15). This difference requires further investigation. It may arise from increased statistical error and insufficient resolution at the highest column densities.

6. For LNPL $N$-pdfs the column density profile has a central PL component with slope -$1/\Gamma$ and an outer LN component which truncates the profile to zero. In most cases it matches a self-gravitating cylinder model. Its central concentration increases as $\sigma_\eta$ increases, as $\Gamma$ decreases, and as the attachment point of the PL component approaches the peak of the $N$-pdf.

7. The "fertility" of a bound filament indicates how much protostar mass can form in the next free-fall time from initial gas within its half-mass radius, compared to that of a uniform filament having the same mass per length and minimal mean column density. The fertility increases with increasing mean column density and with central concentration. For filaments whose $N$-pdfs are dominated by their PL tails, $f$ increases from unity for an initially uniform filament of low mean column density to values greater by an order of magnitude as $\overline{N}$ increases and as $\Gamma$ approaches 1.

8. Characteristic filament mass profiles for the GMCs G216-2.5, Auriga, and NGC3603 have half-mass radii matching those of self-gravitating cylinders with progressively greater central concentration. These half-mass radii have the same values as pure PL models with density exponents $p = 1.3$, 1.4, and 1.7 respectively. The fertility increases from a few for G216-2.5 and Auriga to $f = 18$ for NGC3603. This progression of fertility parallels the incidence of star formation in the three regions.

9. The structure of a characteristic filament may give a useful way to classify the star formation fertility of molecular clouds by their $N$-pdfs. Regions whose $N$-pdfs are power laws of shallow slope are more fertile for star formation, because their more concentrated filaments can more rapidly convert their mass into cores and protostars than can regions of steep $N$-pdf slope.

**Acknowledgements.** Helpful discussions are gratefully acknowledged with João Alves, Philippe André, Blakesley Burkhart, Andi Burkert, Paola Caselli, Hope Chen, Jan Forbricht, Alyssa Goodman, Jouni Kainulainen, Ralf Klessen, Charlie Lada, Stella Offner, Nicola Schneider, and Rowan Smith. Nicola Schnieder also kindly provided estimated uncertainties on $N$-pdf fit parameters. The referee made useful suggestions which improved the paper.